\documentclass[11pt]{amsart}
\reversemarginpar
\newcommand{\commentout}[1]{}

\usepackage{amsmath}
\usepackage{amssymb}

\newcommand{\nwc}{\newcommand}
\nwc{\kvec}{\vec{\bk}}
\nwc{\bkvec}{\vec{\bk}}
\newcommand{\half}{\frac{1}{2}}
\newcommand{\lt}{\left}
\newcommand{\rt}{\right}
\newcommand{\vas}{\varepsilon}
\newcommand{\lan}{\left\langle}
\newcommand{\ran}{\right\rangle}
\newcommand{\tvas}{\Psi_z^\vas}
\newcommand{\psiep}{\Psi_z^\vas}

\newcommand{\veptil}{\tilde{\ml V}_z^\vas}

\newcommand{\cv}{{\ml V}^\ep_z}
\newcommand{\cvtil}{\tilde{{\ml V}}^\ep_z}

\newcommand{\ks}{\tilde{k}}
\newcommand{\bx}{\mathbf x}
\newcommand{\bp}{\mathbf p}
\newcommand{\by}{\mathbf y}

\nwc{\nwt}{\newtheorem}

\nwt{prop}{Proposition}
\nwt{proposition}{Proposition}
\nwt{assumption}{Assumption}
\nwt{lemma}{Lemma}
\nwt{theorem}{Theorem}
\nwt{cor}{Corollary}
\nwt{corollary}{Corollary}
\nwt{remark}{Remark}
\nwt{definition}{Definition} 

\nwc{\bal}{\begin{align}}
\nwc{\be}{\begin{equation}}
\nwc{\ben}{\begin{equation*}}
\nwc{\bea}{\begin{eqnarray}}
\nwc{\beq}{\begin{eqnarray}}
\nwc{\bean}{\begin{eqnarray*}}
\nwc{\beqn}{\begin{eqnarray*}}
\nwc{\beqast}{\begin{eqnarray*}}

\nwc{\eal}{\end{align}}
\nwc{\ee}{\end{equation}}
\nwc{\een}{\end{equation*}}
\nwc{\eea}{\end{eqnarray}}
\nwc{\eeq}{\end{eqnarray}}
\nwc{\eean}{\end{eqnarray*}}
\nwc{\eeqn}{\end{eqnarray*}}
\nwc{\eeqast}{\end{eqnarray*}}

\nwc{\ep}{\varepsilon}
\nwc{\ept}{\epsilon}
\nwc{\vrho}{\varrho}
\nwc{\orho}{\bar\varrho}
\nwc{\ou}{\bar u}
\nwc{\vpsi}{\varpsi}
\nwc{\lamb}{\lambda}

\nwc{\nn}{\nonumber}
\nwc{\bm}{\boldmath}
\nwc{\mf}{\mathbf}
\nwc{\mb}{\mathbf}
\nwc{\ml}{\mathcal}

\nwc{\IA}{\mathbb{A}} 
\nwc{\IB}{\mathbb{B}}
\nwc{\IC}{\mathbb{C}} 
\nwc{\ID}{\mathbb{D}} 
\nwc{\IM}{\mathbb{M}} 
\nwc{\IP}{\mathbb{P}} 
\nwc{\II}{\mathbb{I}} 
\nwc{\IE}{\mathbb{E}} 
\nwc{\IF}{\mathbb{F}} 
\nwc{\IG}{\mathbb{G}} 
\nwc{\IN}{\mathbb{N}} 
\nwc{\IQ}{\mathbb{Q}} 
\nwc{\IR}{\mathbb{R}} 
\nwc{\IT}{\mathbb{T}} 
\nwc{\IZ}{\mathbb{Z}} 

\nwc{\cE}{{\ml E}}
\nwc{\cP}{{\ml P}}
\nwc{\cL}{{\ml L}}
\nwc{\cR}{{\ml R}}
\nwc{\cV}{{\ml V}}
\nwc{\cT}{{\ml T}}
\nwc{\crV}{{\ml V}_{(\delta,\rho)}}
\nwc{\cC}{{\ml C}}
\nwc{\cA}{{\ml A}}
\nwc{\cK}{{\ml K}}
\nwc{\cB}{{\ml B}}
\nwc{\cD}{{\ml D}}
\nwc{\cF}{{\ml F}}
\nwc{\cM}{{\ml M}}
\nwc{\cG}{{\ml G}}
\nwc{\cH}{{\ml H}}
\nwc{\bk}{{\mb k}}

\begin{document}

\title{Scaling Limits for 
 Beam Wave Propagation
in Atmospheric Turbulence}

\author{Albert C. Fannjiang$^*$ \hspace{2cm}
 Knut Solna$^\dagger$}
\thanks{$*$Department of Mathematics,
University of California at Davis,
Davis, CA 95616
Internet: fannjian@math.ucdavis.edu. Research is supported
by The Centennial Fellowship of American Mathematical Society
and UC Davis Chancellor's Fellowship.\\
\hspace{0.5cm}$\dagger$ Department of Mathematics,
University of California at Irvine,
Internet: ksolna@math.uci.edu.
}

\begin{abstract}
We prove the convergence of the solutions of
the parabolic wave equation to that of the Gaussian
white-noise model widely used in the physical
literature. The random medium is isotropic and
is assumed to have integrable correlation coefficient
in the propagation direction. We discuss
the limits of vanishing inner scale and
divergent outer scale of the turbulent medium.
\end{abstract}

\maketitle
\section{Introduction}
The small-scale
refractive index variations, called
the refractive turbulence, in the atmosphere
is the result of small scale fluctuations of
temperature, pressure and humidity caused
by the turbulence of air velocities.
For optical propagation in the atmosphere the influence of
the temperature variations on the refractive
index field is dominant whereas
in the microwave range, the effect of the humidity
variations is more important.
The refractive turbulence results in the
phenomena of beam wander, beam broadening
and intensity fluctuation (scintillation).
It is important to note that these effects depend
on the length scales of the waves as well as
the refractive turbulence \cite{Ma}.

The refractive turbulence  is modeled on the basis of
Kolmogorov theory of turbulence which introduces the notion of
the inertial range bounded by the outer scale $L_0$
(of
the order of $100 {\tt m} - 1{\tt km}$) and the inner
scale $\ell_0$ (of the order of $1-10 {\tt mm}$).
Other features of the refractive turbulence
in the open clear atmosphere include \cite{St}: (i) small changes
(typical value of $3 \times 10^{-4}$ at sea level)
in the refractive index related to
small variations in temperature (on the order
of $0.1-1^o C$), (ii) small scattering angle
which is of the order $\lamb/\ell_0$
and has the typical value $3\times 10^{-4} \mbox{\tt rad}$
for $\lamb=0.6 {\tt mn}$ and $\ell_0=2 {\tt mm}$.
Perturbation methods for solving the Maxwell equations are adequate provided
that the propagation distance is less than, say,
$100 {\tt m}$, a severe limitation on their applicability to
imaging or communication problems.
Our motivation is mainly from laser or microwave beams but
our consideration and results apply equally well
to ultrasound waves in atmospheric turbulence.
The results are also relevant in the context of ultrasound waves
penetrating through complicated multiscale fluctuating (interface)
zones in for instance human tissue.

Under the condition $\lamb=O(\ell_0)$ (including the
millimeter and the sub-millimeter range) the depolarization
term in the Helmholtz equation for the electric field
is negligible \cite{St} and one can use the (scalar)
Helmholtz equation
\beq
\label{helm}
\nabla^2 E+k^2\bar{n}^2(1+\tilde{n})^2 E=0
\eeq
with appropriate boundary conditions
where $k$ is the wavenumber, $\bar{n}$ is
the mean refractive index field and
$\tilde{n}$ is the normalized fluctuation
of the refractive index.

\subsection{The rescaled parabolic approximation}

The well-known parabolic approximation
to equation (\ref{helm}) is  applicable
in a regime where
the variations of the index of refraction are small on
the scale of the wavelength
so that backscattering is negligible \cite{St}. 
This is almost always valid for laser beam in the
atmosphere.

In this paper we study the
initial value problem for
the parabolic wave equation
\beq
\label{par}
\nabla^2_\perp \Psi(z,\bx)+2i k \frac{\partial\Psi(z,\bx)}{
\partial z}=-2k^2\tilde{n}(z,\bx)\Psi(z,\bx),\quad
\Psi(0,\bx)=F_0\left(\frac{\bx}{a}\right) \in L^2(\IR^2)
\eeq
where $ z$ is the longitudinal coordinate in the
direction of the propagation, $\bx=(x_1,x_2)$ is the
transverse coordinates, $\nabla_\perp$ is the
transverse gradient and $\Psi$ is related to
the scalar wave field
$E$ by $E=\Psi(z,\bx)\exp{(ik\bar{n}z)}$. The initial
condition has a typical width $a$ which is the aperture.
Below we will
drop the perp in denoting
the derivatives in the transverse directions.

The difficulty in solving equation (\ref{par}) lies in the random
multiscale nature of $\tilde{n}(z,\bx)$.
First we non-dimensionalize eq. (\ref{par}) as follows.
Let $L_z$ be the
propagation  distance in the longitudinal direction.
Let $\lambda_0$
be the characteristic
wavelength. The corresponding central wavenumber is
$k_0=2\pi/\lambda_0$. The Fresnel length $L_f$ is then
given by
\[
L_f=\sqrt{L_z/k_0}.
\]
We introduce dimensionless wave number and 
coordinates
\[
\ks=k/k_0,\quad \tilde{\bx}=\bx/L_f,\quad  \tilde{z}=z/L_z
\]
 and rewrite the equation in the form 
\beq
2i\ks \frac{\partial \Psi}{\partial z} +
\Delta\Psi+2
  \ks^2 {k_0 L_z} \tilde{n}(z L_z,
{\bx L_f})\Psi=0,\quad\Psi(0,\bx)=F_0(\gamma^{1/2}\bx)\in L^2(\IR^2)
\label{0.3}
\eeq
after dropping
the tilde in the coordinate variables where
\[
\gamma=\lt(\frac{L_f}{a}\rt)^2
\]
is assumed to be $O(1)$,
thus the source is supported on the scale determined by the Fresnel length.

\subsection{Model spectra}

A widely used model for the structure function
of the refractive index field of the atmosphere
is based on the Kolmogorov theory of turbulence
and has the following modified Von K'arm'an spectral density
\beq
\label{0.1}
\Phi_n(\kvec)=0.033C_n^2(|\bkvec|^2+K_0^2)^{-11/6}\exp{(-|\bkvec|^2/K_m^2)}
\eeq
where $\bkvec=(\xi,\bk)$, with $\xi\in \IR, \bk\in \IR^2$
the Fourier variables conjugate to the longitudinal
and transversal coordinates, respectively. Here
$K_0=2\pi/L_0, K_m=5.92/\ell_0$.
This spectrum has the correct behavior
only in the inertial subrange, i.e.
\beq
\label{kol}
\Phi_n(\bkvec)\sim |\bkvec|^{-11/3},\quad |\bkvec|\in (2\pi L_0^{-1},
2\pi \ell_0^{-1}).
\eeq
Outside of this range, particularly for $|\bkvec|\ll
2\pi L_0^{-1}$ there is no physical basis for their
behavior; they are just mathematically convenient expressions
of the cutoffs. In particular, if the wave statistics
strongly depend on $\ell_0$ or $L_0$, then the
problem probably requires more accurate information
on the refractive index field outside of the inertial range
\cite{CHM}, \cite{Fa},
\cite{Fa2}.
Note that the ratio $L_0/\ell_0$
grows like $\mbox{Re}^{3/4}$
as the Reynolds number $\mbox{Re}$
tends to infinity.

There are several variants of (\ref{0.1}) arising from
modeling more detailed features of the refractive index field.
One of
them is 
the Hill spectrum \cite{An}, \cite{Hi} 
to account for the ``bump'' at high wave numbers
which is known to occur near
the inner scale
\beq
\label{hi}
\Phi_n(\bkvec)=0.033 C_n^2\left[1+1.802|\bkvec|/K_m-0.254(|\bkvec|/K_m)^{7/6}\right]
\left(|\bkvec|^2+K_0^2\right)^{-11/6}\exp{(-|\bkvec|^2/K_m^2)}
\eeq
where $K_m=3.3/\ell_0$. 
The coefficient $C_n^2$ is itself a random variable that depends
on time as well as the altitude.
Note that in atmospheric turbulence the inner and outer
scales and the exponent in the power law may also have to be modeled as
stochastic processes \cite{PS}.
The temporal dependence is irrelevant
for optical propagation; the altitude dependence has a rather
permanent, non-universal structure with length scales much greater
than the outer scale $L_0$ \cite{Ma}.

We will consider a class of spectra satisfying the
upper bound
\beq
\label{power'}
{\Phi(H, \bkvec)}
&\leq&
K
(L_0^{-2}+|\bkvec|^2)^{-H-3/2}
\lt(1+\ell_0^2|\bkvec|^2
\rt)^{-2},
\bkvec=(\xi,\bk)\in
\IR^{3},H\in (0,1)
\eeq
with some constant $K<\infty$
as the ratio $L_0/\ell_0\to\infty$ in the high Reynolds number
limit. 
 The details of the spectrum are not pertinent to our
results, only the exponent $H$ is. In particular,
$H=1/3$ for the modified Von K'arm'an spectrum (\ref{0.1}). 

\subsection{White noise scaling}
Let us introduce the non-dimensional parameters that are pertinent to our
scaling:
\[
\ep = \sqrt{\frac{L_f}{L_z}},\quad
\eta=\frac{L_f}{L_0},\quad
\rho=\frac{L_f}{\ell_0}.
\]
In terms of the parameters and the power-law spectrum in
(\ref{power'}) we  rewrite (\ref{0.3}) as
\beq
2i\ks \frac{\partial \Psi^\ep}{\partial z} +
\Delta\Psi^\ep+\frac{\ks^2}{\ep} \sigma_H
\cV(\frac{z}{\ep^2},
\bx)\Psi^\ep=0, \quad\Psi^\ep(0,\bx)=F_0(\gamma^{1/2}\bx)\in L^2(\IR^2).
\label{para}
\eeq
with
\beq
\sigma_H = 
\frac{L_f^{H}}{\ep^3}{\mu}
\eeq
where $\mu$ is the standard deviation of 
the  refractive index field corresponding to 
$\Phi(H,\bkvec)$. 
The spectrum for the (normalized) process $\cV$ is given by
\beq
\label{power} {\Phi_{\eta,\rho}( \bkvec)}
&\leq &K
(\eta^2+|\bkvec|^2)^{-H-3/2}
\lt(1+\rho^{-2}|\bkvec|^2
\rt)^{-2},
\bkvec=(\xi,\bk)\in
\IR^{3}, H\in (0,1)
\eeq
which is rescaled version of (\ref{power'}). 
For high Reynolds number one has
$L_0/\ell_0=\rho/\eta\gg 1$ which is always
the case in our study.

In the beam approximation one has $\ep\ll 1$.
The beam approximation is well within the range
of validity of the parabolic approximation.
The white-noise scaling then corresponds to
$\sigma_H=O(1)$. We set it
to unity by absorbing the constant into
$\cV$. 
This implies relatively weak
fluctuations of the index field, i.e.
\[
\tilde{C_n}\sim L_f^{3/2-H}L_z^{-3/2}
\ll 1,\quad\hbox{as}\,\,L_z\to\infty
\]
in view of the fact that $ H\in (0,1)$ and $\ep\ll 1$.

In the present paper we first study  the white-noise
scaling with $\rho<\infty$ and $\eta>0$ fixed
as $\ep\to 0$. We then discuss the resulting
white-noise model with $\rho\to \infty$ and
$\eta\to 0$. 
For the proof, we adopt the approach of \cite{rich} where
the turbulent transport of passive scalars is studied.
In \cite{whn-wig} the white noise limit is studied via the so
called Wigner distribution.

\section{Formulation and main results}
\subsection{Martingale formulation}
We consider the
weak formulation  of the equation:
\beq
 i\ks\lt[\lan \Psi_z^\varepsilon, \theta\ran - \lan \Psi_0,
\theta \ran\rt] &=&
-\int_0^z \half\lan \Psi_s^\vas, \Delta \theta\ran ds
-\frac{\ks^2}{\vas}\int_0^z \lan \Psi_s^\vas, \cV(\frac{s}{\vas^2},\cdot)\cdot
\theta
\ran ds
\label{weak}
\eeq
for any test function $\theta \in C_c^\infty(\IR^2)$,
the space of smooth functions with compact support.
The tightness result (Section~4.1) implies that
for $L^2$ initial data 
the limiting measure $\IP$ is supported
in the Skorohod space $D([0,z_0];L_w^2(\IR^2))$.
Here and below $L^2_w(\IR^2)$ denotes the standard $L^2$-function
space with the weak topology.

For tightness as well as identification of the limit,
 the following infinitesimal operator  $\cA^\ep$ will play an important role.
Let $\cV^\vas_z\equiv \cV(z/\ep^2,\cdot)$, $\mathcal{F}_z^\vas$ the
$\sigma$-algebras generated by $\{\cV_s^\vas, \, s\leq z\}$  and
$\mathbb{E}_z^\vas$ the corresponding conditional expectation w.r.t. $\cF^\ep_z$.
Let $\cM^\ep$ be the space of measurable functions adapted to $\{\cF^\ep_z, \forall z\}$ 
such that $\sup_{z<z_0}\IE|f(z)|<\infty$.
We say $f(\cdot)\in \cD(\cA^\ep)$, the domain of $\cA^\ep$, and $\cA^\ep f=g$
if $f,g\in \cM^\ep$ and for
$f^\delta(z)\equiv\delta^{-1}[\IE^\ep_z f(z+\delta)-f(z)]$
we have
\beqn
\sup_{z,\delta}\IE|f^\delta(z)|&<&\infty\\
\lim_{\delta\to 0}\IE|f^\delta(z)-g(z)|&=&0,\quad\forall z.
\eeqn
Consider the special class of admissible functions
 $f(z)=\phi(\lan \Psi_z^\vas, \theta\ran),
 f'(z)=\phi'(\lan \Psi_z^\vas, \theta\ran),
 \forall \phi\in C^\infty(\IR)$,
then we have the following expression
from (\ref{weak}) and the chain rule 
\beq
\label{gen}
 \cA^\vas
f(z)
&=&  if'(z)\lt[\frac{1}{2\ks} \lan \Psi_z^\vas,
\Delta \theta\ran + \frac{\ks}{\vas} \lan \Psi_z^\vas,
\cv\theta\ran\rt].
\eeq
 A main property of $\cA^\ep$ is
that 
\beq
\label{12.2}
f(z)-\int^z_0 \cA^\ep f(s) ds\quad\hbox{is a  $\cF^\ep_z$-martingale},
\quad\forall f\in \cD(\cA^\ep).
\eeq
Also,
\beq
\label{mart}
\IE^\ep_s f(z)-f(s)=
\int^z_s \IE^\ep_s \cA^\ep f(\tau) d\tau\quad \forall s<z \quad\hbox{a.s.}
\eeq
(see \cite{Kur}).
We denote by $\cA$ the infinitesimal operator corresponding
to the unscaled process $\cV_z(\cdot)=\cV(z,\cdot)$.

Define
\beq
\label{gamma1}
\Gamma^{(1)}(\bx,\by)
&=&\int\int\int^\infty_0\cos{((\bx-\by)\cdot \bp)}\cos{(s\xi)}
\Phi_{(\eta,{\rho})}(\xi,\bp)\,\,ds\,d\xi\,d\bp\\
&=&\pi\int\cos{((\bx-\by)\cdot
\bp)}\Phi_{(\eta,{\rho})}(0,\bp)\,\,d\bp
\nn
\\
{\Gamma}^{(1)}_0(\bx)&=&{\Gamma}^{(1)}(\bx,\bx)
\eeq
where we have written the wavevector $\bkvec\in \IR^3$ as
$\bkvec=(\xi,\bp)$ with $\bp\in\IR^2$.

Now we formulate the solutions for the Gaussian Markovian model
as the solutions to the corresponding martingale problem:
Find a measure $\IP$ (of $\Psi_z$) on
the subspace of $D([0,\infty);L^2_{w}(\IR^2))$
whose elements have the initial condition $F_0(\gamma^{1/2}\bx)$
such that
 \beqn
&&f(\lan \Psi_z,\theta\ran)-\int_0^z
 \bigg\{f'( \lan \Psi_s,\theta\ran)\left[\frac{i}{2\ks}\lan
 \Psi_s,\Delta\theta\ran-{\ks^2}
 \lan \Psi_s,{\Gamma}^{(1)}_0\theta\ran\right]
  -{\ks^2}f''(\lan \Psi_s,\theta\ran) 
  \lan\theta, {\cK}_{\Psi_s}\theta\ran\bigg\}\,ds\\
\nonumber&&\hbox{{\em is a martingale w.r.t. the filtration of a cylindrical
Wiener process, for each} $f\in C^\infty(\IR)$}
\eeqn
where
\beq
\label{r10}
 {\cK}_{\Psi_s}\theta
 =\int \Psi_s(\bx)\Psi_s(\by){\Gamma}^{(1)}(\bx,\by)
 \theta(\by)
 \,d\by.
 \eeq
 The Gaussian Markovian model has been extensively studied for
 beam wander, broadening and scintillation effects
 in the literature (see, e.g. \cite{Ba}, \cite{FPS}).
 It can also been written as 
 the It\^{o}'s equation
 \beq
 d\Psi_z&=&
 \left(\frac{i}{2\ks}\Delta-
 {\ks^2}{\Gamma}^{(1)}_0\right)
\Psi_z\,dz+{i\ks}\left({\cK}_{\Psi_z}\right)^{1/2}
 \,dW_z\nn\\
 &=&\left(\frac{i}{2\ks}\Delta -
 {\ks^2}{\Gamma}^{(1)}_0\right)
 \Psi_z\,dz+{i\ks}\Psi_z
d\tilde{W}_z\nn\\
 &=&\frac{i}{2\ks}\Delta
 \Psi_z\,dz+{i\ks}\Psi_z \circ
d\tilde{W}_z,\quad\Psi_0(\bx)=
 F_0(\gamma^{1/2}\bx)
\label{ito}
 \eeq
  where $\circ$ stands for the Stratonovich
integral, and ${W}_z(\bx)$ and
$\tilde{W}_z(\bx)$ are the Brownian fields with the spatial
  covariance $\delta(\bx-\by)$ and ${\Gamma}^{(1)}(\bx,\by)$,
respectively. 

The existence and uniqueness for the Schr\"{o}dinger-It\^o
eq.
(\ref{ito})  with $\rho<\infty$ and $\eta>0$ has been studied
in
\cite{DP} by using the Wiener chaos expansion.
Note that  the following limit exists
\beq
\label{gamma2}
\bar{\Gamma}(\bx,\by)
&=&\lim_{\rho\to\infty}\pi\int\cos{((\bx-\by)\cdot
\bp)}\Phi_{(\eta,{\rho})}(0,\bp)\,\,d\bp\\
&=&\pi\int\cos{((\bx-\by)\cdot
\bp)}\Phi_{(\eta,\infty)}(0,\bp)\,\,d\bp,\quad\eta>0
\nn
\eeq
By the well-posedness result of \cite{DP}
and a standard
weak$-\star$ compactness argument one can prove the existence
of weak solution in
$D([0,\infty); L^2_w(\IR^2))$ for $\rho=\infty, H\in (0,1)$.

Next we consider the limiting case $\eta= 0$.  This 
 would
induce uncontrollable large scale fluctuation  
the
 Gaussian, Markovian model which
should be factored out first. 
 Thus we consider
the solution of
the form
\[
\Psi(z,\bx)={\Psi}'(z,\bx)\exp{(
{i\ks}\int^z_0\tilde{W}_s(0)\,ds)}
\]
and the resulting equation
\beq d\Psi_z
&=&\frac{i}{2\ks}\Delta
 \Psi_z\,dz+{i\ks}\Psi_z \circ
d\tilde{W}'_z,\quad\Psi_0(\bx)=
 F_0(\gamma^{1/2}\bx)
\label{ito2}
\eeq
where $\tilde{W}_z'$ is given by
  \beq
   \label{1.6}
    \tilde{W}_z'(\bx)&=&\tilde{W}_z(\bx)-\tilde{W}_z(0)
\eeq
with the covariance function
    \beq
    \label{large}
   \overline{\Gamma'}(\bx,\by)
   &=&\pi\int (e^{i\bx\cdot\bp}-1)(e^{-i\by\cdot\bp}-1)
    \Phi_{(0,\infty)}(0,\bp)d\bp.
    \nn
    \eeq
    Note that the above integral  is convergent
    only if
    \[
    H<1/2;
    \]
    in particular, the limit exists
    for the modified Von K'arm'an spectrum $H=1/3$.
Since $H<1/2$, the limiting model is only
H\"{o}lder continuous in the transverse coordinates.

Again by the well-posedness result of \cite{DP}
and a standard
weak$-\star$ compactness argument one can prove the existence
of weak solution in
$D([0,\infty); L^2_w(\IR^2))$ for 
$\rho=\infty, \eta=0, H\in (0,1/2)$.

\subsection{Uniqueness}
		   Because of the non-smoothness (when $\rho=\infty$) and  
the non-homogeneity (when $\eta=0$)
of the white-noise
		  potential  in the transverse coordinates the
uniqueness argument of
\cite{DP} does not apply
								here.

Taking the  function $f(r)=r^n$ in the martingale
formulation, we arrive after some algebra at the following
equation
\beq
\label{npt}
\frac{\partial F_z^{(n)}}{\partial z}&=&
\cC_1  F_z^{(n)} +\cC_2  F_z^{(n)}
\eeq
\nn
for the $n-$point correlation function
\[
F_z^{(n)}(\bx_1,\dots,\bx_n)\equiv
\IE\lt[\Psi_z(\bx_1)\cdots\Psi_z(\bx_n)\rt]
\]
where
\begin{eqnarray}
\cC_1 &=&\frac{i}{2\ks}\sum_{j=1}^{n}\Delta_{\bx_j}\\
\label{25}
\cC_2&=&
-{\ks^2}\sum_{{j,k=1}}^{n}\overline{\Gamma}
(\bx_j,\bx_k),\\
\label{26}
\hbox{or} ~~  \cC_2  &=&
-{\ks^2}\sum_{{j,k=1}}^{n}\overline{\Gamma}'
(\bx_j,\bx_k)\\
\end{eqnarray}
We will now establish the uniqueness for 
eq. (\ref{npt}) with the initial data
\[
F_0^{(n)}(\bx_1,\dots,\bx_n)=
\IE\lt[\Psi_0(\bx_1)\cdots\Psi_0(\bx_n)\rt],\quad\Psi_0\in L^2(\IR^2).
\]

In the former case (\ref{25}) 
 $\cC_2$ is 
a {\em bounded}, H\"older continuous function and we rewrite
eq. (\ref{npt}) in the mild formulation
\beq
\nn
 F_z^{(n)}=\exp{(z\cC_1)}
 F_0^{(n)}+
 \int^z_0\exp{[(z-s)\cC_1]}
 \cC_2 F_s^{(n)}\,\,ds
 \eeq
 whose local existence and uniqueness can be easily established
 by straightforward application of the contraction mapping principle.
 By linearity, local well-posedness can be extended to
 global well-posedness.

In the latter case (\ref{26}) $\cC_2$
 is unbounded, H\"older continuous function
with sub-Lipschitz growth. We first note that
$\cC_2$ is  non-positive everywhere
since
\[
\sum_{{j,k=1}}^{n}\overline{\Gamma}'
(\bx_j,\bx_k)=
\pi\int \sum_{j}(e^{i\bx_j\cdot\bp}-1)\overline{\sum_{k}(e^{i\bx_k\cdot\bp}-1)}
    \Phi_{(0,\infty)}(\bp)d\bp\geq 0.
    \]
Hence both $\cC_1$ and $\cC_2$ 
  are generators of one-parameter
  contraction semigroups on $L^2(\IR^{2n})$,  thus by
  the product formula (Theorem~3.30, \cite{Da}) we have
  \[
  \lim_{m\to\infty}\lt[\exp{(\frac{z}{m}\cC_1)}
  \exp{(\frac{z}{m}\cC_2)
     }\rt]^m F=\exp{[z(\cC_1+\cC_2)]}F
     \]
     for all $F\in L^2(\IR^{2n})$, which then gives
     rise to a unique semigroup on $L^2(\IR^{2n})$.

\subsection{Main assumptions and theorem}
Let $V_z$ be a $z$-stationary, $\bx$-homogeneous 
square-integrable process whose
spectral density satisfies the upper bound
(\ref{power}).

Let $\cF_z $ and $\cF^+_z$  be the sigma-algebras generated by
$\{V_s:  \forall s\leq z\}$ and $\{V_s: \forall s\geq  z\}$,
respectively. Define
the correlation coefficient
\beq
\label{correl}
\rho(t)=\sup_{h\in \cF_z\atop \IE[h]=0,
\IE[h^2]=1}
\sup_{
g\in \cF_{z+t}^+\atop \IE[g]=0,
\IE[g^2]=1}\IE\lt[h g\rt].
\eeq

\begin{assumption}
The correlation coefficient $\rho(t)$ is integrable
\end{assumption}

When $V_z$ is a Gaussian process, the correlation coefficient
$\rho(t)$ equals the {\em linear} correlation coefficient  $r(t)$
which has the following useful  expression
\beq
\label{corr}
r(t) &=&\sup_{g_1, g_2}
 \int R(t-\tau_1-\tau_2,\bk) g_1(\tau_1,\bk)
g_2(\tau_2,\bk)d\bk d\tau_1 d\tau_2
\eeq
where
\[
R(t,\bk)=\int e^{it\xi} \Phi(\xi, \bk) d\xi
\]
and the
supremum is taken over all $g_1, g_2\in L^2(\IR^{d+1})$ which are
supported on $(-\infty, 0]\times
\IR^d$ and  satisfy the constraint
\beq
\label{b.1}
\int R(t-t',\bk) g_1(t,\bk)\bar{g}_1(t',\bk) dt dt' d\bk=
\int R(t-t',\bk) g_2(t,\bk)\bar{g}_2(t',\bk) dt dt' d\bk=1.
\eeq
There are various criteria for the decay rate (e.g.,
exponential decay) of the
linear correlation coefficients, see \cite{IR}.

\begin{lemma}
\label{lemma-new}
Assumption~1 implies that 
the random field
\[
\tilde{\cV}_z(\bx)=\int^\infty_0 \IE_z[\cV_t(\bx)]dt
\]
is $\bx$-homogeneous and has a finite second moment which
satisfies the upper bound:
\beqn
\IE[\tilde{\cV}_z^2] &\leq &
\int^\infty_z \int^\infty_z
\lt|\IE\lt[\IE_z[\cV_t]\IE_z\lt[\cV_s\rt]\rt]\rt|ds dt \\
&\leq&
\IE[\cV_z^2]\lt(\int^\infty_0\rho(t)dt\rt)^2.
\eeqn
\end{lemma}
 \begin{proof}
Consider in the definition of the correlation coefficient
\beqn
h_1=&\IE_z(V_s) &\in L^2(\Omega,P, \cF_z)\\ 
h_2= &V_t &\in L^2(\Omega,P,\cF^+_t).
\eeqn
We then  have
\beqn
\lt|\IE\lt[ \IE_z[\cV_s(\bx)]\IE_z[\cV_t(\bx)]\rt]\rt|=
\lt|\IE\lt[\IE_z[\cV_s(\bx)] \cV_t(\bx)\rt]\rt|
\leq \rho(t-z) \IE^{1/2}\lt[\IE_z^2[\cV_s]\rt]
\IE^{1/2}\lt[V^2_t\rt].
\eeqn
Hence by  setting $s=t$ first and the Cauchy-Schwartz
inequality we have
\beqn
\IE\lt[\IE^2_z[\cV_s]\rt]&\leq&\rho^2(s-z)\IE[\cV_z^2]\\
\lt|\IE\lt[ \IE_z[\cV_s(\bx)]\IE_z[\cV_t(\bx)]\rt]\rt|
&\leq& \rho(t-z) \rho(s-z) \IE[\cV_z^2],\quad s, t \geq z.
\eeqn
Therefore
\beq
\IE[\tilde{\cV}_z^2] &\leq &
\int^\infty_z \int^\infty_z
\lt|\IE\lt[\IE_z[\cV_t]\IE_z\lt[\cV_s\rt]\rt]\rt|ds dt \nn\\
&\leq&
\IE[\cV_z^2]\lt(\int^\infty_0\rho(t)dt\rt)^2\label{n.1}
\eeq
which, together with the integrability of $\rho(t)$,  implies a finite second
order moment of
$\tilde{\cV}_z$. 

\end{proof}

\begin{corollary}
\label{corr-new} For each $L, z_0<\infty$
and $\rho<\infty, \eta>0$
there exists a constant $\tilde{C}$ such that
\[
\sup_{z<z_0\atop |\bx|\leq L}
\IE\lt[\Delta \tilde{\cV}_{\lamb z}\rt]^2
\leq \tilde{C}
\]
for all $H\in(0,1),\lamb\geq 1$.
\end{corollary}
\begin{proof}
Analogous to (\ref{n.1}) we have
\beqn
\IE[\Delta\tilde{\cV}_z]^2
&\leq&
\IE[\Delta \cV_z]^2\lt(\int^\infty_0\rho(t)dt\rt)^2.
\eeqn
A straightforward spectral calculation shows
that 
\[
\IE[\Delta \cV_z(\bx)]^2=
O(\rho^{4-2H}),\quad\forall \bx, z.
\]

\end{proof}

It is easy to see that
\[
\cA\tilde{\cV}_z=-\cV_z
\]
and that
\[
\Gamma^{(1)}(\bx,\by)=\IE\lt[\tilde{\cV}_z(\bx)\cV_z(\by)\rt]
\]
where $\Gamma^{(1)}$ is given by (\ref{gamma1}).

Next we assume the following quasi-Gaussian property:
\begin{assumption}
\beq
\label{ga1}
\sup_{|\by|\leq L}\IE\lt[ V^\ep_z(\by)
\rt]^4&\leq& C_1  \sup_{|\by|\leq L|}\IE^2\lt[
V^\ep_z\rt]^2(\by)\\
\label{ga2}
\sup_{|\by|\leq L}\IE\lt[\tilde{\cV}^\ep_z
\rt]^4(\by)&\leq& C_2\sup_{|\by|\leq L} \IE^2\lt[
\tilde{\cV}^\ep_z\rt]^2(\by)\\
\label{ga3}
\sup_{|\by|\leq L}\IE\lt[\lt[
{V}^\ep_z \rt]^2\lt[
\tilde{\cV}^\ep_z \rt]^4\rt](\by)&\leq& C_3\lt\{\lt(\sup_{|\by|\leq L}
\IE\lt[ V^\ep_z\rt]^2(\by)\rt)\lt(
\sup_{|\by|\leq L} \IE^2\lt[
\tilde{\cV}^\ep_z\rt]^2(\by)\rt)\rt.\\
&&\lt.+
\lt(\sup_{|\by|\leq L}
\IE^2\lt[ V^\ep_z\tilde{V}^\ep_z\rt](\by)\rt)\lt(
\sup_{|\by|\leq L} \IE\lt[
\tilde{\cV}^\ep_z\rt]^2(\by)\rt)
\rt\}
\nn
\eeq
for all $L<\infty$ where the constants $C_1, C_2$ and $C_3$ are independent
of $\ep, \eta, \rho, \gamma$. 
\end{assumption}

\begin{assumption}
For any fixed $\eta>0$ and 
every $\theta\in C^\infty_c(\IR^2)$
\beq
\label{1.4'}
\sup_{z<z_0}\|\theta\tilde{\cV}(\frac{z}{\ep^2},\cdot)\|_2
 & = & o\lt(\frac{1}{\ep}\rt),\quad 
\forall \ep\leq  1\leq \rho
 \eeq
with
a random constant of finite moments independent of
$\rho$ and $\ep$.
\end{assumption}

When $\cV$ is {\em Gaussian}, $\tilde{\cV}$ is
also Gaussian and
  condition (\ref{1.4'})
  is  always satisfied
  \beq
  \sup_{z<z_0}\|\theta\tilde{\cV}(\frac{z}{\ep^2},\cdot)\|_2
  &\leq&\tilde{ C}\log \left[\frac{z_0}{\vas^2} \right]
\eeq
where the random constant $\tilde{C}$ has a Gaussian-like tail
by a simple application of Borell's inequality \cite{Ad}.

\begin{theorem}
\label{thm1}
Let $\cV$ satisfy  Assumptions 1, 2 and 3. Let 
$\eta>0$ and $\rho<\infty$  be fixed  as
$\ep\to 0$.
Then
the weak solution $\Psi^\ep$ of (\ref{weak}) converges in
the space of $D([0,\infty);L_w^2(\IR^2))$
to that of
the Gaussian white-noise model with the covariance functions
${\Gamma}^{(1)}$ and ${\Gamma}^{(1)}_0$.
\end{theorem}

Note that in the limiting model with $\rho=\infty$ the
white-noise velocity field has transverse regularity of
H\"{o}lder exponent $H+1/2$.

The convergence of the white-noise limit
has been established in \cite{BCF} and \cite{BF}
for a refractive index field that is a function of
$z$ only. 

\section{Proof of Theorem~1}
\subsection{Tightness}

In the sequel we will adopt the following notation 
\[
f(z)\equiv f(\lan \Psi_z^\vas, \theta\ran),\quad
 f'(z)\equiv f'(\lan \Psi_z^\vas, \theta\ran),\quad
f''(z)\equiv f''(\lan \Psi_z^\vas, \theta\ran),\quad 
\quad\forall f\in C^\infty(\IR).
\]
Namely, the prime stands for the differentiation w.r.t. the original argument (not $z$)
of $f, f'$ etc.

A family 
of processes $\{\Psi^\ep, 0<\ep<1\} \subset D([0,\infty);L^2_{w}(\IR^2)) 
$ is 
tight if and only if the family of 
processes
$\{\lan \Psi^\ep, \theta\ran, 0<\ep <1\}
\subset D([0,\infty);L^2_{w}(\IR^2))
$ is tight for all $\theta\in C^\infty_c(\IR^2)$. 
We use the tightness  criterion of \cite{Ku}
(Chap. 3, Theorem 4), namely, 
we will prove:
Firstly,
\beq
\label{trunc}
& & \lim_{N\to \infty}\limsup_{\ep\to 0}\IP\{\sup_{z<z_0}|\lan \Psi_z^\ep, \theta\ran|
\geq N\}=0,\quad\forall z_0<\infty.
\eeq
Secondly, for  each $f\in C^\infty(\IR)$  
there is a sequence
$f^\ep(z)\in\cD(\cA^\ep)$ such that for each $z_0<\infty$
$\{\cA^\ep f^\ep (z), 0<\ep<1,0<z<z_0\}$ 
is uniformly integrable and
\beq
\label{19}
& & \lim_{\ep\to 0} \IP\{\sup_{z<z_0} |f^\ep(z)-
f(\lan \Psi^\ep, \theta\ran) |\geq \delta\}=0,\quad \forall \delta>0.
\eeq
Then it follows that the laws of
$\{\lan \Psi^\ep, \theta\ran, 0<\ep <1\}$ are tight in the space
of $D([0,\infty);L^2_{w}(\IR^2))$

Condition (\ref{trunc}) is satisfied because the $L^2$-norm is
preserved.
Let
\[
f_1^\vas (z)\equiv \frac{i\ks}{\vas}\int_z^\infty
\mathbb{E}_z^\vas\, f'(z) \lan \Psi_z^\vas,\cV^\ep_s\theta\ran\,ds
\]
be the 1-st perturbation of $f(z)$. 
Let
\[
\tilde{\cV}^\ep_z=\frac{1}{\ep^2}\int^\infty_z \IE_z^\ep \cV^\ep_s\,ds.
\]
We obtain
\begin{equation}
\label{1st}
f_1^\vas (z)= 
{i\ks\vas}
f'(z) \lan \Psi_z^\vas,\cvtil\theta\ran.
\end{equation}

\begin{prop}\label{prop:2}
\begin{enumerate}
$$\lim_{\ep\to 0}\sup_{z<z_0} \mathbb{E} |f_1^\vas(z)|=0,\quad
\lim_{\ep\to 0}\sup_{z<z_0} |f_1^\vas(z)|= 0
\quad \hbox{in probability}$$.
\end{enumerate}
\end{prop}

\begin{proof}
We have
\beq
\label{1.2}
& & \mathbb{E}[|f_1^\vas(z)|]\leq {\vas}\|f'\|_\infty
\|\Psi_0\|_2 
\IE\|\theta\tilde{\cV}^\ep_z\|_2
\eeq
and
\beq
\label{1.3}
& & \sup_{z< z_0} |f_1^\vas(z)|
 \leq {\vas}
\|f'\|_\infty  \|\Psi_0\|_2
\sup_{z<z_0}\|\theta\cvtil\|_2.
\eeq
The right side of (\ref{1.2}) is $O(\ep)$  by
Lemma~\ref{lemma-new} while that of (\ref{1.3}) is $o(1)$ in
probability by Assumption~3.

\end{proof}

Set $f^\ep(z)=f(z)+f^\ep_1(z)$.
A straightforward calculation yields
\beqn
 \cA^\vas f_1^\vas &=&-{\vas}f''(z)\lt[\lan
\Psi_z^\vas,\Delta\theta \ran
+\frac{\ks^2}{\ep} 
\lan \Psi_z^\vas,\cv\theta\ran\rt]\lan\psiep,\cvtil\theta\ran\\
&&\quad -{\ep}f'(z)\lt[\half\lan
 \Psi_z^\vas,\Delta(\cvtil\theta)\ran
+
\frac{\ks^2}{\ep}\lan
\Psi_z^\vas,\cv\cvtil\theta\ran\rt]-\frac{i\ks}{\ep}
f'(z)\lan\psiep,\cv\theta\ran
\nn
\eeqn
and, hence
\beq
\label{25'}
\cA^\vas f^\ep(z)
&=&\frac{i}{2\ks}f'(z)\lan \tvas,\Delta\theta \ran -
{\ks^2}f'(z)\lan\psiep,\cv\cvtil\theta\ran
-{\ks^2}f''(z)\lan\psiep,\cv\theta\ran\lan\psiep,\cvtil\theta\ran\\
&&-\frac{\ep}{2}\lt[ f'(z)\lan
\psiep,\Delta(\cvtil\theta)\ran
+f''(z)\lan\psiep,\Delta\theta\ran\lan\psiep,\cvtil\theta\ran
\rt]\nonumber \\
&=&A_1^\vas(z)+A_2^\vas(z)+A_3^\vas(z)+A_4^\vas(z)
\nonumber
\eeq
where $A_2^\vas(z)$ and $A_3^\vas(z)$ are the $O(1)$ statistical coupling
terms.

For the tightness criterion stated in the beginnings of the section,
it remains to show
\begin{prop}
\label{lemma2}
$\{\cA^\ep f^\ep\}$ are uniformly integrable and
$$\lim_{\ep\to 0}\sup_{z<z_0}\IE|A^\ep_4(z)|=0$$.
\end{prop}

\begin{proof}
We show that $\{A^\ep_i\}, i=1,2,3,4$ are uniformly integrable. 
To see this, we have the following estimates.
\bean
|A_1^\vas(z)|
&\leq&\frac{1}{2\ks}\|f'\|_\infty\|\Psi_0\|_2\|\Delta \theta\|_2\\
|A_2^\vas(z)|
&\leq&{\ks^2}\|f'\|_\infty\|\Psi_0\|_2
\|\cv\cvtil\theta\|_2\\
|A_3^\vas(z)|
&\leq&{\ks^2}\|f''\|_\infty\|\Psi_0\|^2_2
\|\cv\theta\|_2\|\cvtil\theta\|_2.
\eeqn
For fixed $\eta$, the second moments of the right hand side
of the above expressions are uniformly bounded as
$\ep\to 0$
and hence $A_1^\vas(z), A_2^\vas(z), A_3^\vas(z)$ are uniformly integrable.
 \beq
 |A^\ep_4|
  \nonumber
  &\leq& \frac{\vas}{2}
  \left[\|f''\|_\infty\|\Psi_0\|^2_2
  \|\Delta\theta\|_2\|\cvtil\theta\|_2+
  \|f'\|_\infty\|\psiep\|_2\|\Delta(\cvtil\theta)\|_2\rt].
 \label{1.10}
  \eeq
  By Lemma~\ref{lemma-new} and Corollary~\ref{corr-new}
   $A^\ep_4$ is  uniformly integrable.
  Finally, it is clear that  $$\lim_{\ep\to 0}\sup_{z<z_0}\IE|A^\ep_4(z)|=0.$$
  \end{proof}

\subsection{Identification of the limit}
Once the tightness is established we can use another result 
 in \cite{Ku} (Chapter 3, Theorem 2) to identify the limit.
The setting there is finite-dimensional but
the argument is entirely  applicable to the
infinite-dimensional setting here (cf. \cite{whn-wig}).

Let $\bar{\cA}$ be a diffusion or jump diffusion operator such that
there is a unique solution $\omega_z$ in the space 
$D([0,\infty);L^2_{w}(\IR^2))$
 such that 
\beq
\label{38}
f(\omega_z)-\int^z_0\bar{\cA} f(\omega_s)\,ds
\eeq
is a martingale. 
We shall show that for each $f\in C^\infty(\IR)$ there exists $f^\ep\in \cD({\cA}^\ep)$
such that
\beq
\label{38.2}
\sup_{z<z_0,\ep}\IE|f^\ep(z)-f(\lan \Psi^\ep_z,\theta\ran)|&<&\infty\\
\label{39.2}
\lim_{\ep\to 0}\IE|f^\ep(z)-f(\lan \Psi^\ep_z,\theta\ran)|&=&0,\quad \forall z<z_0\\
\label{40.2}
\sup_{z<z_0,\ep}\IE|{\cA}^\ep f^\ep(z)-\bar{\cA} f(\lan \Psi^\ep_z,\theta\ran)|&<&\infty\\
\lim_{\ep\to 0}\IE|{\cA}^\ep f^\ep(z)-\bar{\cA} f(\lan \Psi^\ep_z,\theta\ran)|&=&0,\quad
\forall z<z_0.
\label{42.2}
\eeq
Then it follows that any tight processes
$\lan \Psi^\ep_z,\theta\ran$ converges
in law to the unique process generated by $\bar{\cA} $.
As before we adopt the notation $f(z)=f(\lan \Psi^\ep_z,\theta\ran)$.

For this purpose,
we introduce the next perturbations $f_2^\ep, f_3^\ep$.
Let
\bea
\label{40.3}
A_2^{(1)}(\phi) &\equiv&\int\int\theta(\bx)\phi(\bx)
\Gamma^{(1)}(\bx,\by)\phi(\by)\theta(\by)\,\,d\bx\,d\by\\
A_3^{(1)}(\phi)&\equiv&\int\Gamma^{(1)}(\bx,\bx)\phi(\bx)\theta(\bx)\,\,d\bx
\label{41.2}
\eea
where 
\beq
    {\Gamma}^{(1)}(\bx,\by)&\equiv&
    \IE\lt[\cv(\bx)\cvtil(\by)\rt].
\eeq
It is easy to see that
\beq
A_2^{(1)}(\phi)=\mathbb{E}\lt[\lan\phi, \cv\theta\ran
\lan\phi, \cvtil\theta\ran\rt].
\eeq

Define
\begin{align*}
f_2^\vas(z) &\equiv
{\ks^2}f''(z)\int_z^\infty \mathbb{E}_z^\vas
\lt[\lan\tvas, \cV^\ep_s\theta\ran\lan\tvas,\tilde{\cV}^\ep_s\theta\ran -A^{(1)}_2(\tvas)\rt]\,ds
\\
f_3^\vas(z) &\equiv {\ks^2}f'(z)\int_z^\infty
\mathbb{E}_z^\vas
\lt[\lan\tvas, \cV^\ep_s(\tilde{\cV}^\ep_s\theta)\ran-A^{(1)}_3(\tvas)\rt]\,ds.
\end{align*}
Let
\[
\Gamma^{(2)}(\bx,\by)\equiv
\IE\lt[\veptil(\bx)\veptil(\by)\rt],
\]
and
\bea
A_2^{(2)}(\phi) &\equiv&\int\int\theta(\bx)\phi(\bx)
\Gamma^{(2)}(\bx,\by)\phi(\by)\theta(\by)\,\,d\bx\,d\by\\
A_3^{(2)}(\phi)&\equiv&\int\Gamma^{(2)}(\bx,\bx)\phi(\bx)\theta(\bx)\,\,d\bx,
\eea
we then have
\beq
f_2^\vas(z) &=&\frac{\ep^2\ks^2}{2}f''(z)
\lt[\lan\psiep,\cvtil\theta\ran^2-A^{(2)}_2(\psiep)\rt]\\
f_3^\vas(z) &=&
\frac{\ep^2\ks^2}{2}f'(z)\lt[\lan\psiep,\cvtil\cvtil\theta\ran-
A^{(2)}_3(\psiep)\rt].
\eeq

\begin{prop}\label{prop:4}
$$ \lim_{\ep\to 0}\sup_{z<z_0} \mathbb{E}|f_2^\vas(z)|=0,\quad \lim_{\ep\to 0}\sup_{z<z_0}
\mathbb{E}|f_3^\vas(z)|=0. $$
\end{prop}
\begin{proof}
We have the bounds
\beqn
\sup_{z<z_0}\IE|f_2^\vas(z)|&\leq&
\sup_{z<z_0}{\vas^2\ks^2}\|f''\|_\infty\lt[
\|\Psi_0\|_2^2\IE\|\cvtil\theta\|_2^2
+\IE[A^{(2)}_2(\psiep)]\rt]\\
\sup_{z<z_0}
\IE|f_3^\vas(z)|&\leq&
\sup_{z<z_0}
{\vas^2\ks^2}\|f'\|_\infty
\lt[\|\Psi_0\|_2\IE \|\cvtil\cvtil\theta\|_2
+\IE[A_3^{(2)}(\psiep)]\rt];
\eeqn
both of them tend to zero.
\end{proof}

We have
\beqn
\cA^\vas f_2^\vas(z)&=&{\ks^2}
f''(z)\left[-
\lan\tvas, \cv\theta\ran\lan\tvas,\cvtil\theta\ran + A^{(1)}_2(\tvas)\right]
+ R_2^\vas(z)\\
\cA^\vas f_3^\vas(z)&=&{\ks^2}
f'(z)\left[-\lan
\tvas,\cv(\cvtil\theta)\ran + A^{(1)}_3(\tvas)\right]+
R_3^\vas(z)
\eeqn
with
\beq
\nn
R_2^\vas(z)& =&{i\ep^2\ks}\frac{f'''(z)}{2}\left[\half
\lan\psiep,\Delta\theta\ran+\frac{\ks^2}{\ep}\lan\psiep,\cv\theta\ran\rt]
\lt[\lan\psiep,\cvtil\theta\ran^2-A_2^{(2)}(\psiep)\rt]\\
\nn
&& +
{i\ep^2\ks}f''(z)\lan\psiep,\cvtil\theta\ran\lt[
\half\lan\psiep,\Delta(\cvtil\theta)\ran+\frac{\ks^2}{\ep}
\lan\psiep,\cv\cvtil\theta\ran\rt]\\
&&-{i\ep^2\ks}f''(z)\lt[\half\lan\psiep,\Delta(G_\theta^{(2)}\psiep)\ran+
\frac{\ks^2}{\ep}\lan\psiep,\cv G_\theta^{(2)}\psiep\ran\rt]
\label{36}
\eeq
where 
$G_\theta^{(2)}$ denotes the operator
\[
G_\theta^{(2)}\phi\equiv \int \theta(\bx)\Gamma^{(2)}(\bx,\by)\theta(\by)\phi(\by)
\,d\by.
\]
Similarly
\beqn
R^\ep_3(z)&=&{i\ep^2\ks}
f'(z)\left[\half\lan\psiep,\Delta(\cvtil\cvtil\theta)\ran+\frac{\ks^2}{\ep}
\lan\psiep,\cv\cvtil\cvtil\theta\ran\rt]\\
&&+{i\ep^2\ks}
f''(z)\left[\half
\lan\psiep,\Delta\theta\ran+\frac{\ks^2}{\ep}\lan\psiep,\cv\theta\ran\rt]
\lt[\lan\psiep,\cvtil\cvtil\theta\ran-A_3^{(2)}(\psiep)\rt]\\
&&-{i\ep^2\ks}f'(z)\lt[\half
\lan\psiep,\Delta(\Gamma^{(2)}_0\theta)\ran+\frac{\ks^2}{\ep}
\lan\psiep,\cv\Gamma^{(2)}_0\theta\ran\rt]
\eeqn
where
\[
\Gamma^{(2)}_0(\bx)\equiv\Gamma^{(2)}(\bx,\bx).
\]

\begin{prop}\label{prop:5}
\[
\lim_{\ep\to 0}\sup_{z<z_0} \mathbb{E} |R_2^\vas(z)|=0,\quad \lim_{\ep \to 0}
\sup_{z<z_0} \mathbb{E}
|R_3^\vas(z)|=0.
\]
\end{prop}
The argument is entirely analogous to that for Proposition~\ref{prop:4}.
The most severe factors involve $\Delta(\cvtil \theta)$ and 
$\Delta(\cvtil\cvtil \theta)$, both of which have uniformly
bounded second moments by Assumption~2 and
Corollary~\ref{corr-new}. Therefore the corresponding terms
are $O(\ep^2)$.

Consider the test function $f^\ep(z)=f(z)+f_1^\ep(z)
-f_2^\ep(z)-f_3^\ep(z)$. We have
\beq
\label{2.67}
\lefteqn{\cA^\ep f^\ep(z)}\\
&=&
\frac{i}{2\ks}f'(z)\lan\psiep,\Delta\theta\ran-{\ks^2}f''(z)
A_2^{(1)}(\psiep)-{\ks^2}f'A_3^{(1)}(\psiep)
-R_2^\ep(z)-R_3^\ep(z)+A_4^\ep(z).\nn
\eeq
Set
\beq
\label{remainder}
R^\vas(z) = R_1^\vas(z) - R_2^\vas(z) - R_3^\vas(z),\quad\hbox{with}\,\,
R_1^\vas(z)=A_4^\ep(z).
\eeq
It follows from Propositions 3 and 5 that
\[
\lim_{\ep \to 0}\sup_{z<z_0}\IE|R^\ep(z)|=0.
\]
Recall that
\beqn
M_z^\vas(\theta)&=&f^\ep(z)-\int^z_0 \cA^\ep f^\ep(s)\,ds\\
&=& f(z)+f_1^\vas(z)-f_2^\vas(z)-f_3^\vas(z)
-
\int_0^z\frac{i}{2\ks}f'(z)\lan\tvas,\Delta\theta\ran\,ds\\
&& + \int_0^z{\ks^2}\left[f''(s)
A_2^{(1)}(\Psi_s^\vas)+f'(s) A_3^{(1)}(\Psi_s^\vas)\right]\,ds -\int_0^z
R^\vas(s)\,ds
\eeqn
is a martingale. Now that (\ref{38.2})-(\ref{42.2}) are satisfied
we can identify the limiting martingale to be
\begin{equation}
M_z(\theta)=f(z)-\int_0^z
\bigg\{f'(s)\left[\frac{i}{2\ks}\lan
\Psi_s,\Delta\theta\ran-{\ks^2}{A}^{(1)}_3(\Psi_s)\right]
 - {\ks^2}f''(s){A}^{(1)}_2(\Psi_s)\bigg\}\,ds.
 \label{37}
\end{equation}

Since $\lan\tvas,\theta\ran$ is uniformly bounded
\[
\lt|\lan\tvas,\theta\ran\rt|\leq \|\Psi_0\|_2{\|\theta\|}_2
\]
we have the convergence of the second moment
\[
\lim_{\ep\to 0}
\IE\left\{
{\lan\tvas,\theta\ran}^2\right\}=\mathbb{E}\left\{ {\lan
\Psi_z,\theta\ran}^2\right\}.
\]
Use $f(r) =r$ and $r^2$ in (\ref{37})
$$ M_z^{(1)}(\theta)=\lan \Psi_z,\theta\ran -
\int_0^z \left[\frac{i}{2\ks}\lan \Psi_s,\Delta\theta\ran -
{\ks^2}{A}^{(1)}_3(\Psi_s)\right]\,ds $$ is a
martingale with the quadratic variation
\[
\left[M^{(1)}(\theta),M^{(1)}(\theta)\right]_z=
{-\ks^2}\int_0^z{A}^{(1)}_2(\Psi_s)\,ds
={-\ks^2}\int^z_0\lan\theta,
{\cK}_{\Psi_s}\theta\ran\,ds
\]
where
\[
{\cK}_{\Psi_s}\theta=
\int\Psi_s(\bx)\Gamma^{(1)}(\bx,\by)\Psi_s(\by)\theta(\by)\,d\by.
\]
Therefore, $$
M_z^{(1)}={i\ks}
\int_0^z \sqrt{{\cK}_{\Psi_s}}dW_s $$ where
$W_s$ is a real-valued, cylindrical Wiener process (i.e.
$dW_z(\bx)$ is a space-time white noise field)
and $\sqrt{{\cK}_{\Psi_s}}$ is the square-root of
the positive-definite operator given in (\ref{r10}).

\end{document}